\documentclass[twocolumn,pra,showpacs,floatfix,superscriptaddress]{revtex4}
\usepackage{amsmath}
\usepackage{amsfonts}
\usepackage{amsthm}
\usepackage{graphicx}
\usepackage{amssymb}
\usepackage[matrix,frame,arrow]{xy}
\usepackage{amsmath}
\newcommand{\bra}[1]{\left\langle{#1}\right\vert}
\newcommand{\ket}[1]{\left\vert{#1}\right\rangle}
\newcommand{\qw}[1][-1]{\ar @{-} [0,#1]}
\newcommand{\qwx}[1][-1]{\ar @{-} [#1,0]}
\newcommand{\gate}[1]{*+[F]{#1} \qw}
\newcommand{\control}{*-=-{\bullet}}
\newcommand{\ctrl}[1]{\control \qwx[#1] \qw}
\newcommand{\targ}{*{\xy{<0em,0em>*{} \ar @{ - } +<.4em,0em> \ar @{ - } -<.4em,0em> \ar @{ - } +<0em,.4em> \ar @{ - } -<0em,.4em>},*+<.8em>\frm{o}\endxy} \qw}
\newcommand{\multigate}[2]{*+{\hphantom{#2}} \qw \POS[0,0].[#1,0] !C *{#2} \POS[0,0].[#1,0] \drop\frm{-}}
\newcommand{\ghost}[1]{*+{\hphantom{#1}} \qw}
\newcommand{\gategroup}[6]{\POS"#1,#2"."#3,#2"."#1,#4"."#3,#4"!C*+<#5>\frm{#6}}
\newcommand{\rstick}[1]{*!L!<-.5em,0em>=<0em>{#1}}
\newcommand{\lstick}[1]{*!R!<.5em,0em>=<0em>{#1}}
\newcommand{\dstick}[1]{*!U!<0em,.5em>=<0em>{#1}}
\newcommand{\Qcircuit}{\xymatrix @*=<0em>}
\newcommand{\nuc}[2]{\mbox{${}^{#1}\rm #2$}}
\newcommand{\identity}{\ensuremath{\mathbf{1}}}
\newcommand{\units}[2]{\mbox{$#1\,\text{#2}$}}
\newcommand{\para}{\textit{para}}
\newcommand{\Para}{\textit{Para}}
\newcommand{\pHH}{\para-hydrogen}
\newcommand{\PHH}{\Para-hydrogen}

\newcommand{\HH}{\mbox{$\text{H}_2$}}
\newcommand{\hydride}{\mbox{$\text{Ru(H)}_2\text{(CO)}_2\text{(dppe)}$}}

\newcommand{\ashydride}{\mbox{$\text{Ru(H)}_2\text{(CO)}_2\text{(dpae)}$}}
\newcommand{\asprecursor}{\mbox{$\text{Ru(CO)}_3\text{(dpae)}$}}
\newcommand{\asintermediate}{\mbox{$\text{Ru(CO)}_2\text{(dpae)}$}}

\begin{document}
\title{Implementing Grover's Quantum Search on a \PHH\ based Pure State NMR Quantum Computer}
\author{M.~S. Anwar}
\email{muhammad.anwar@physics.ox.ac.uk} \affiliation{Centre for
Quantum Computation, Clarendon Laboratory, University of Oxford,
Parks Road, OX1 3PU, United Kingdom}

\author{D.~Blazina}
\email{db30@york.ac.uk} \affiliation{Department of Chemistry,
University of York, Heslington, York, YO10 5DD, United Kingdom}

\author{H.~A. Carteret}
\email{cartereh@iro.umontreal.ca} \affiliation{LITQ, Departement
d'Informatique et Recherche Op\'erationelle, Pavillon
Andr\'e-Aisenstadt, Universit\'e de Montr\'eal, Montr\'eal,
Qu\'ebec H3C 3J7, Canada}

\author{S.~B. Duckett}
\email{sbd3@york.ac.uk} \affiliation{Department of Chemistry,
University of York, Heslington, York, YO10 5DD, United Kingdom}

\author{J.~A. Jones}
\email{jonathan.jones@qubit.org} \affiliation{Centre for Quantum
Computation, Clarendon Laboratory, University of Oxford, Parks
Road, OX1 3PU, United Kingdom}

\date{\today}

\begin{abstract}
We demonstrate the implementation of Grover's quantum search
algorithm on a liquid state nuclear magnetic resonance (NMR)
quantum computer using essentially pure states.  This was achieved
using a two qubit device where the initial state is an essentially
pure ($\varepsilon=1.06\pm0.04$) singlet nuclear spin state of a
pair of \nuc{1}{H} nuclei arising from a chemical reaction
involving \pHH. We have implemented Grover's search to find one of
four inputs which satisfies a function.
\end{abstract}
\maketitle

% main text
\section{Introduction} \label{Intro}
Quantum computers \cite{BennetDiv} are of enormous interest
because of their ability to perform calculations which appear to
be intractable on any conceivable classical computer. The leading
technology for the implementation of quantum logic gates is liquid
state nuclear magnetic resonance (NMR)
\cite{ernst87,levitt,freeman}, and NMR quantum computers
\cite{cory96,cory97,jones98a,chuang98a} have been used to
implement a range of quantum algorithms, mostly notably Shor's
algorithm which was used to factor fifteen \cite{vandersypen01}.
Unfortunately NMR devices have one great disadvantage: the small
energy gap between the Zeeman levels that are used as
computational basis states means that direct cooling is not a
practical method for obtaining a pure initial state.  The most
common way round this is to prepare a pseudopure state
\cite{cory96}, whose behaviour is identical to that of a pure
state up to a scaling factor, but this is not a good solution for
two reasons. Firstly, the efficiency of pseudopure state
preparation falls off exponentially with the number of qubits in
the quantum computer \cite{warren97}, which means that NMR devices
cannot be scaled up to useful sizes, and secondly the states used
are provably separable\cite{braunstein99}, and so cannot exhibit
the phenomenon of entanglement.

These problems are not inherent in NMR quantum computation, but
arise from the use of pseudopure states prepared from high
temperature thermal states.  We have recently shown how high
purity initial states can be prepared using \pHH\ techniques
\cite{anwar04a,anwar04c}, and that these states can be used to
perform the simplest quantum computation, Deutsch's algorithm
\cite{anwar04b}.  Similar results have been obtained by H\"ubler
\textit{et al.} \cite{Bargon}, albeit with a lower spin state
purity. Here we describe the use of a new molecular system which
allows essentially pure initial states to be generated, and the
implementation of Grover's quantum search algorithm
\cite{grover97} on our pure state computer. State initialization
on demand is achieved by using a \units{12}{ns} laser pulse to
initiate a rapid chemical reaction, and is therefore both well
controlled and time coherent with the NMR radiofrequency pulses.

\section{\PHH} \label{PHH}
We prepare pure spin states using an effect called \pHH\ induced
polarization (PHIP)
\cite{bowers86,natterer97,duckett99,duckett03}.  The existence of
the \para\ isomer of dihydrogen, \HH, is a consequence of the
Pauli principle, which requires the overall wavefunction to be
antisymmetric with respect to exchange of the fermionic \nuc{1}{H}
nuclei.  It follows that \HH\ molecules in even rotational states,
most notably the $\text{J}=0$ ground state, possess an
antisymmetric nuclear spin wave function and correspond to nuclear
spin singlets, termed \para.  Thus if \HH\ is cooled to a
temperature of \units{20}{K} in the presence of a paramagnetic
catalyst then essentially pure \pHH\ will be obtained, with the
singlet spin wavefunction
\begin{equation}\label{singlet-vector}
|\psi^-\rangle=\,\frac{1}{\sqrt{2}}(\ket{01}-\ket{10}),
\end{equation}
where we have used the computational basis in which $\ket{0}$
corresponds to the ground state and $\ket{1}$ to the excited state
of a spin-1/2 particle.  The \pHH\ molecule cannot be used
directly for NMR quantum computing, due to its high symmetry, but
this can be overcome by using a chemical reaction to prepare a new
molecule, in which the two hydrogen atoms can be made distinct and
can be separately addressed.  For further details see
\cite{anwar04a,anwar04c,anwar04b,Bargon,bowers86,natterer97,duckett99,duckett03}.

In our previous work we have used the two hydride \nuc{1}{H}
nuclei in \hydride, where dppe indicates
1,2-bis(diphenyl\-phosphino)ethane, as our NMR quantum computer.
Detailed analysis of this system shows that the spins are not
prepared in an absolutely pure singlet state, but in a slightly
mixed state, which can be easily converted to a Werner state
\cite{werner89} of the form
\begin{equation}\label{initialstate}
\rho_\textit{init}=(1-\varepsilon)\frac{\identity}{4}+\varepsilon\mbox{$\ket{\psi^-}$}\mbox{$\bra{\psi^-}$},
\end{equation}
with a spin-state purity of $\varepsilon=0.898\pm0.026$
\cite{anwar04c}. Although this system is adequate for an initial
proof of principle experiment, it is not ideally suited for
quantum computing; in particular its \nuc{1}{H} spin--spin
relaxation time ($\text{T}_2=\units{0.58}{s}$) is uncomfortably
short.  We have, therefore, prepared a novel but closely related
system, \ashydride, where dpae indicates
1,2-bis(diphenyl\-arsino)ethane. This system has a slightly longer
\nuc{1}{H} spin--spin relaxation time
($\text{T}_2=\units{0.67}{s}$), and has the further advantage that
the initial state can be prepared with a purity indistinguishable
from one (the purity was measured using techniques described
previously \cite{anwar04a,anwar04c} and was found to be
$\varepsilon=1.06\pm0.04$). Thus we describe our new system as a
pure state NMR quantum computer.

\begin{figure}
\begin{center}
\mbox{ \Qcircuit @C=1em @R=1em {
\dstick{\ket{\psi^-}} & & & \ctrl{1} & \gate{H} & \targ & \qw & \ket{0}\\
                      & & & \targ    & \qw      & \targ & \qw & \ket{0}
\gategroup{1}{3}{2}{4}{1em}{\{}} }
\end{center}
\caption{Quantum circuit to convert a singlet state $\ket{\psi^-}$
into the conventional initial state $\ket{00}$.  $H$ indicates a
single qubit Hadamard gate, $\oplus$ indicates a single qubit
\textsc{not} gate, and the initial two qubit gate is a
controlled-\textsc{not} gate.} \label{fig:disentangle}
\end{figure}
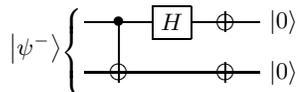

Although our quantum computer starts in a pure state, it does not
start in the state conventionally assumed for two qubit quantum
computers, that is $\ket{00}$ in the computational basis.  It is,
however, simple to convert our initial state to this form by using
a disentangling circuit, such as that shown in
Fig.~\ref{fig:disentangle}

\section{Grover's quantum search}\label{Grover}
Grover's quantum search algorithm \cite{grover97} permits an
efficient search for one of $k$ matching items within a search
space of size $n$. The algorithm is most conveniently described in
terms of a binary function, with $n$ possible inputs of which $k$
\textit{satisfy} the function, that is produce an output of $1$.
The function is assumed to be implemented by means of an
\textit{oracle}, which will produce values of the function for any
given input, but does not permit any other method of analyzing the
function.  The best classical algorithm is simply to try inputs at
random, and this will allow a satisfying input to be located in
$O(n/k)$ queries. By contrast, Grover's quantum search allows a
satisfying input to be located in $O(\sqrt{n/k})$ queries.  The
simplest case occurs when $k=1$ and $n=4$, in which case the
satisfying input can be located in a single query, and this is the
case we will concentrate on here.

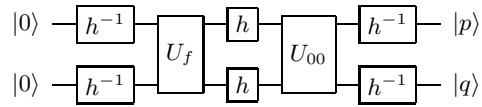
\begin{figure}
\begin{center}
\mbox{ \Qcircuit @C=1em @R=1em @!R {
\lstick{\ket{0}} & \gate{h^{-1}} & \multigate{1}{U_{f}} & \gate{h} & \multigate{1}{U_{00}} & \gate{h^{-1}} & \rstick{\ket{p}} \qw \\
\lstick{\ket{0}} & \gate{h^{-1}} &        \ghost{U_{f}} & \gate{h}
&        \ghost{U_{00}} & \gate{h^{-1}} & \rstick{\ket{q}} \qw } }
\end{center}
\caption{A quantum circuit to implement Grover's search algorithm
on a two qubit quantum computer.  Following previous practice
\cite{jones98} Hadamard gates $H$ have been replaced by
pseudo-Hadamard gates $h$ and $h^{-1}$, which correspond to
$90^\circ_{\pm y}$ rotations. The function being studied is
encoded in the propagator $U_f$, and $U_{00}$ replaces $\ket{00}$
by $-\ket{00}$ while leaving other basis states unchanged.  At the
end of the computation the two qubits are left in states
corresponding to the satisfying input, that is
$f(pq)=1$.}\label{fig:grover}
\end{figure}

The key element in an implementation of Grover's search is a
quantum function evaluation oracle, $U_f$, which performs the
transformation \cite{jones98}
\begin{equation}\label{oracle}
\ket{x}\overset{U_f}{\longrightarrow}(-1)^{f(x)}\ket{x}
\end{equation}
where $\ket{x}$ indicates a quantum register in a state
corresponding to the input $x$, and the function $f$ has values of
$0$ or $1$. Typically $x$ is written in binary form, and $\ket{x}$
is a set of qubits with values corresponding to successive bits of
$x$. A quantum circuit which can be used to locate one satisfying
input from four using this oracle is shown in
Fig.~\ref{fig:grover}.

\section{The Experiment}\label{experiment}
Our two qubit system comprises the hydride \nuc{1}{H} nuclei in
\ashydride, where the hydride atoms are derived from \pHH.  The
precursor compound \asprecursor\ was prepared from
$\text{Ru}_3(\text{CO})_{12}$ and dpae using techniques described
previously \cite{anwar04a,schott02}.  Essentially pure \pHH\ was
prepared at a temperature of \units{18}{K} using a charcoal-based
catalyst, and was introduced into a \units{5}{mm} NMR tube
containing \asprecursor\ dissolved in $\text{d}_6$-benzene. The
NMR tube was then transferred into a \units{400}{MHz} spectrometer
fitted with a \nuc{1}{H}/\nuc{31}{P} probe modified for \textit{in
situ} photolysis \cite{godard02}. The spectrometer triggered a
\units{12}{ns} UV pulse of wavelength \units{308}{nm} from an MPB
Technologies MSX-250 pulsed XeCl excimer laser, irradiating the
active region of the NMR sample and producing the unstable species
\asintermediate.  Intermediates of this type have been shown to
react with hydrogen on the sub-microsecond timescale
\cite{cronin95}, which would in this case lead to the formation of
\ashydride.  For further details see
\cite{anwar04a,anwar04c,anwar04b}.

The hydride resonances appear at \units{-7.61}{ppm} and
\units{-7.21}{ppm}, with a frequency separation of
$\delta=\units{160}{Hz}$, while the hydride $J$ coupling
(${}^2J_\text{HH}$) was \units{4.8}{Hz}. The \nuc{1}{H}
transmitter frequency was placed exactly between the two resonance
frequencies.  The laser flash acts as an initialisation switch,
generating the pure state $\ket{\psi^-}$ on demand, which is
subsequently used for the implementation.

The singlet state $\ket{\psi^-}$ can be converted to the desired
initial state $\ket{00}$ using the NMR pulse sequence
\begin{equation}
P_\textit{prep}\equiv[\frac{1}{4\delta}] \; 90_y \; [\frac{1}{4J}]
\; 180_x \; [\frac{1}{4J}] \; 180_y \; [\frac{1}{2\delta}] \; 90_x
\end{equation}
which has been described previously \cite{anwar04b}.  This pulse
sequence is composed of a series of hard RF pulses (which excite
both spins equally) and periods of evolution under the background
Hamiltonian, indicated by enclosing the appropriate evolution time
in brackets.  Following NMR conventions operations are applied
from left to right. No selective pulses are used, but at two
points equal and opposite $z$ rotations are applied to the two
qubits. These were achieved by evolution under the Zeeman
Hamiltonian at frequencies of \units{\pm\delta/2}{Hz}, with
evolution under the small $J$ coupling neglected over these short
periods.  NMR pulse sequences to implement the four possible
functions $f$ with a single satisfying value can be developed in
the same way and are shown below, where each function is
identified by listing its satisfying input.
\begin{equation}
\begin{split}
P_{00} &\equiv [\frac{1}{4J}] \; 90_{-x} \; 90_y \; 90_{-x} \; [\frac{1}{4J}] \; 180_x\\
P_{01} &\equiv [\frac{1}{4J}] \; 180_x \; [\frac{1}{4J}] \; [\frac{1}{2\delta}] \; 180_x\\
P_{10} &\equiv [\frac{1}{2\delta}] \; [\frac{1}{4J}] \; 180_x \; [\frac{1}{4J}] \; 180_x\\
P_{11} &\equiv [\frac{1}{4J}] \; 180_x \; [\frac{1}{4J}] \;
90_{-x} \; 90_y \; 90_{-x}
\end{split}\label{eq:Pf}
\end{equation}
Two of these functions require equal $z$ rotations to be applied
to the two qubits, and these were achieved using composite $z$
pulses.

The final element used in our pulse sequences is a gradient pulse,
indicated by \textit{crush}.  This is a short period of evolution
during which time the magnetic field is temporarily made spatially
inhomogeneous, so that spins in different parts of the sample
experience different Larmor frequencies. These crush pulses were
applied immediately before and after the Grover circuit, at which
points the qubits should be in eigenstates of the computational
basis. These states do not evolve under magnetic fields, and
should not be affected by the crush pulse, but most error terms
which can arise will be dephased \cite{jones98}.

\begin{figure}
\begin{center}
\includegraphics[scale=0.8]{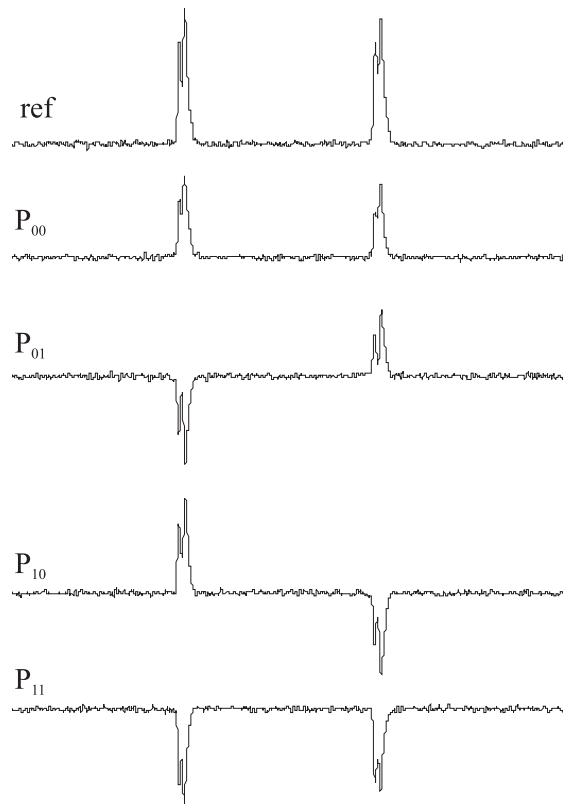}
\end{center}
\caption{Experimental spectra from our implementation of Grover's
quantum search.  The top spectrum is a phase reference acquired
from our computer in the state $\ket{00}$.  The four lower spectra
were acquired after implementing Grover's search with the oracle
set to each of the four functions in turn.} \label{fig:spectra}
\end{figure}

Putting these elements together gives the final pulse sequence
used to implement Grover's search,
\begin{equation}
P_\textit{prep} \; \textit{crush} \; 90_{-y} \; P_f \; 90_y \;
P_{00} \; 90_{-y} \; \textit{crush} \; 90_y \; \textit{acquire}
\end{equation}
where $P_f$ corresponds to one of the four pulse sequences shown
in equation~\ref{eq:Pf}.  The state of the spin system is analyzed
by applying a hard $90^\circ$ pulse and observing the resulting
NMR spectrum. With appropriate phasing the state of each qubit is
then indicated by multiplets pointing upwards for $\ket{0}$ or
downwards for $\ket{1}$.  A phase reference spectrum is easily
obtained using the sequence
\begin{equation}
P_\textit{prep} \; \textit{crush} \; 90_y \; \textit{acquire}
\end{equation}
and the experimental results are shown in Fig.~\ref{fig:spectra}.

These spectra show exactly the pattern expected: each computation
ends with the system in one of the four basis states, and the
final state corresponds with the satisfying input of the
corresponding function.  Note that in this paper we have used the
same conventions as in \cite{anwar04b}, where the first qubit
occurs on the right hand side of the NMR spectrum; this is the
reverse of the convention used in \cite{jones98}. There are small
imperfections visible in the spectra: in particular there are
imbalances in the intensities of the two multiplets and of the two
lines in each multiplet, and the overall signal intensity is lower
in spectra obtained from quantum computations than in the phase
reference spectrum.  The imbalances can be ascribed to errors in
the implementations of the logic gates, while the signal loss is
largely a consequence of relaxation during the pulse sequences.
The effects of relaxation are less marked in this study than in
our previous implementation of Deutsch's algorithm
\cite{anwar04b}; this partly reflects the longer relaxation times
in our new spin system, but is also a consequence of the symmetry
in the gates applied to each qubit in Grover's algorithm.

These results confirm that it is possible to use \pHH\ techniques
to implement quantum algorithms on a pure state NMR quantum
computer.  We are now seeking to extend these results to larger
spin systems and more complex algorithms.

\section*{Acknowledgements}
We thank the EPSRC for financial support.  MSA thanks the Rhodes
Trust for a Rhodes Scholarship.  HAC thanks MITACS for financial
support.  Quantum circuits were drawn using Q-circuit
\cite{qcircuit}.

% \appendix

% \section{}
% \label{}

\end{document}